# Electrical Characterisation of Ultra-thin SAM Structures

*Céline Trapes, Lamia Rouaï*
*Lionel Patrone(\*)*
LACSC, ECE, 53 rue de Grenelle, 75007 Paris
(\*) L2MP-UMR CNRS 6137, ISEN, Place George Pompidou, 83000 Toulon

**ABSTRACT**

The way of reduction of metal oxyde semiconductor (MOS) structures is going to reach limitations and new devices have to be explored as an alternative to MOS technology. Molecular electronic and more particularly self-assembly-molecular technique on silicon substrate gives interesting results as seen in the literature. We are going to study n-alkyltrichlorosilane grafting on oxidised silicon, characterise it macroscopically with ellipsometer and goniometry measurements, and down to microscopic scale with atomic force microscopy. Once the uniformity of the monolayer is verified (roughness of few Angströms) we have tested a sputtering method deposition to form aluminium dots onto the surface. Also metal-insulator-semiconductor diodes are tested measuring both leakage current between gate and substrate and capacitance-voltage. The sputtering method deposition can be improved in order to decrease the gate leakage current and we would like to test another evaporation method. Further application we want to study is gas sensors using conjugated organic films or synthetic polymers and concerns the drift current with gas absorption.

## 1. INTRODUCTION

Miniaturisation and downscaling physical phenomenon study impose to control nano-structures dimension. Lithographic techniques are expensive and a promising solution is the use of self-assembled monolayers (SAM) such as n-alkyltrichlorosilanes on oxidized silicon [1].
This work is an investigation using SAMs to develop organic insulators [2]. Single p-type (100) silicon (Si) is modified by grafting organic molecules onto its surface by using wet chemistry methods. Formation of self-assembled monolayers (SAMs) is verified using ellipsometry to determine the thickness and Atomic Force Microscopy (AFM) to analyse the surface topography. Metal-SAM-Si and Metal-Si (Schottky) samples are tested as resistive diode using current-voltage or capacity-voltage measurements.

## 2. MAIN RESULTS

The SAM formed in our result is a carbonate monolayer with octadecytrichlorosilane (OTS) [$SiCl_3$-$(CH_2)_{17}$-$(CH_3)$] deposited by adsorption [1] on a Silicon wafer which is $10^{-15} cm^{-3}$ p-type doped. The full detail of the deposition is given in the literature [4]. The OTS thickness is evaluated to 2.5 nm with ellipsometry measurement which corresponds to the typical result [3]. Atomic force microscopy observations have confirmed that the OTS is free of pinholes and the uniformity of the monolayer as the roughness is around 2-3 Å rms onto the silicon surface (Fig. 1).

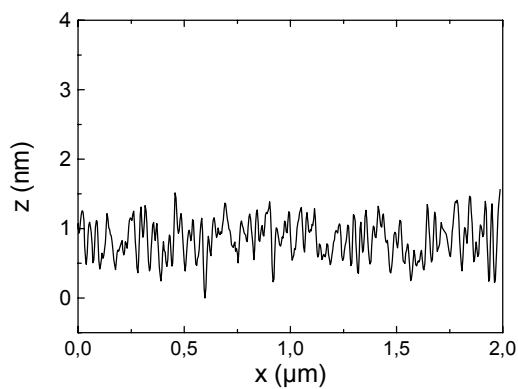

Fig. 1 AFM analysis for a 2µm/2µm picture.

In a second step aluminium dot areas are sputtered to form Metal-OTS-Si diode, dot surface= $1,3\ 10^{-3}\ cm^{-2}$, and dot thickness= 250nm for 20mn long deposition. This OTS insulator formed on a native $SiO_2$ native oxide with Al dot is referred as SAM diode (Fig.3).
We can compare electrical measurements of Schottky diode (Fig.2), for which native oxide is formed on silicon substrate, with SAM diode (Fig.3).

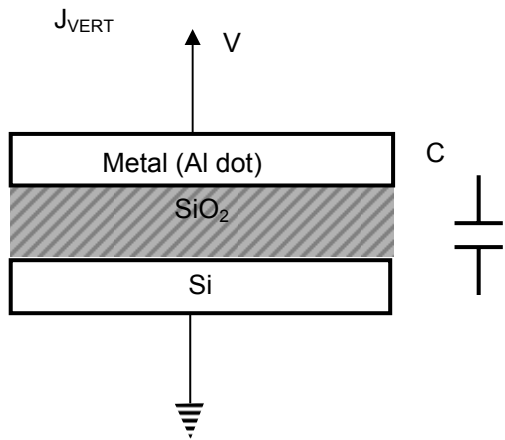

Fig.2: Schottky diode schematic structure. $J_{VERT}$ measurement principle and macro model capacitance C.

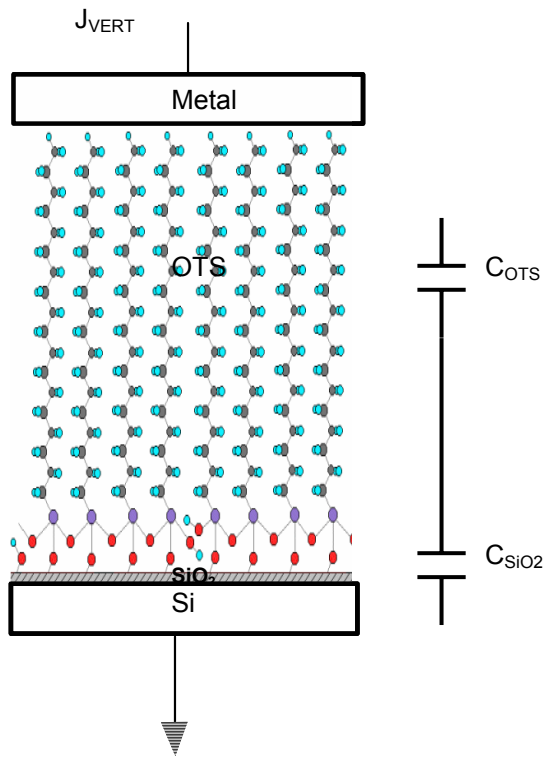

Fig.3: SAM diode schematic structure. $J_{VERT}$ measurement principle and macro model serial capacitance $C_{OTS}$ and $C_{SIO2}$.

Measuring current-voltage characteristics between metal and substrate (vertical measurement "$J_{VERT}$"), we obtained a forward SAM gate current decreased by two orders of magnitude (Fig. 4). In forward condition, the leakage current is about 4 mA/cm² for an average insulating field of 6 MV/cm. In reference to the literature, lower leakage current could be measured thanks to evaporation method deposition which is less damaging for the SAM [2]. Otherwise in a sensor way of research the drift of the I-V characteristics with gas injection is more important than the I-V measure itself.

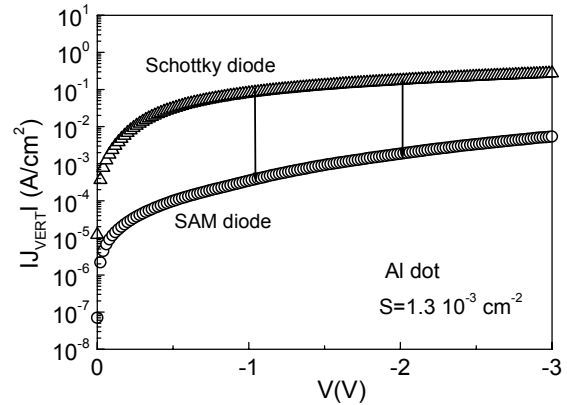

Fig. 4 : Density current measurement in accumulation regime (negative voltage) – comparison of Schottky diode and SAM diode.

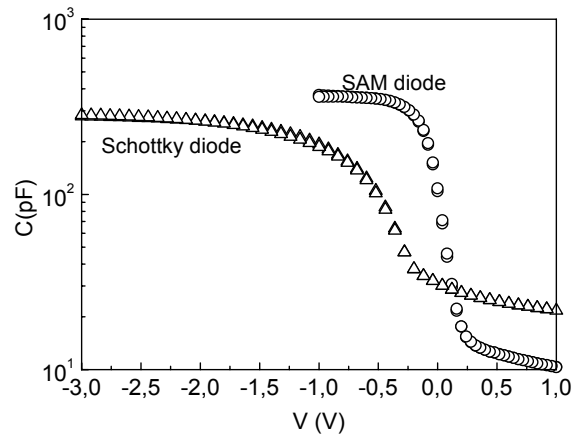

Fig. 5: 800kHz C-V characteristics – sweep rate 5mV/s of Schottky and SAM diode.

Furthermore we have been able to measure the capacitance "C-V" on monolayer as thin as 2.7nm (SAM) down to 1.7nm (Schottky) (Fig. 5). The first difficulty was to choose the best frequency to obtain the expected accumulation region (at negative voltage) characterised by a plateau and this was pretty good at 800 kHz.
Then C-V measurements allow extracting parameters as the dielectric constant ($\varepsilon_{OTS}$) or the monolayer thickness ($T_{OTS}$) thanks to a macro-model based on the serial composition of two capacitances $C_{OTS}$ and $C_{SiO2}$ (Fig.3). Considering this assumption the total measured capacitance C can be calculated as:

$$C = (1/C_{OTS} + 1/C_{SiO2})^{-1}$$

$$C_{OTS} = \frac{\varepsilon_o \cdot \varepsilon_{OTS} \cdot S}{T_{OTS}}$$

Where $\varepsilon_o$ is the vacuum permittivity and S the Al deposited dot surface.

By the way the total capacitance measured at 800 kHz is lower than the expected value: taking dielectrics constants of 3.9 for $SiO_2$ and 2.5 for OTS, then thicknesses of 1.7 nm for $SiO_2$ and 2.7 nm for OTS we calculate that SAM capacitance should be 750pF.

That is why we had to correct the previous model by an additional serial resistance which represents parasitic effects as leakage current. The corrected SAM capacitance (C_corr) extracted in Fig.6 is 650pf which allows extracting an 822pF OTS capacitance against 1nF calculated.

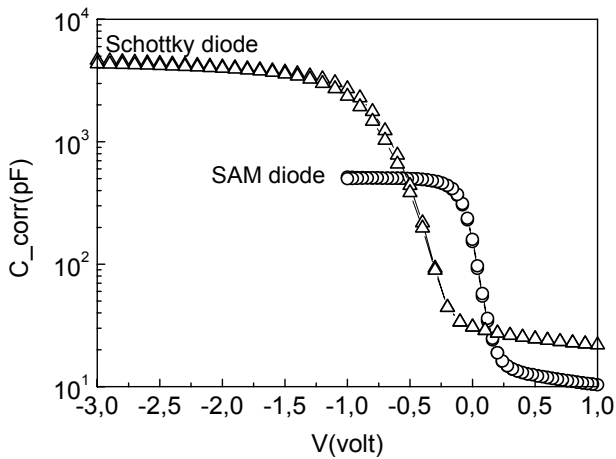

Fig. 6: C-V corrected characteristics –of Schottky and SAM diodes .

## 3. PERSPECTIVES

One application of SAM diodes would be chemical sensors as already tested in the literature [6]. Current-voltage characteristics measured at different gas compositions should involve a modification of the current-voltage measurements due to the junction barrier height modification [7]. These studies would perform relevant investigations of SAM techniques in the area of gas sensors. This experimental study is going to be extended to organic conjugated molecules ended with phenyl groups in order to detect explosive gas as Methane.

## 4. ACKNOWLEDGMENT

Many thanks to the "Laboratoire Matériaux Microélectronique de Provence" (L2MP) for their supports in samples preparation and characterization.